\documentclass[11pt]{article}
\usepackage{amssymb}
\def\be{\begin{equation}}
\def\ee{\end{equation}}
\def\bea{\begin{eqnarray}}
\def\eea{\end{eqnarray}}

\def\y{\'{\i}}

\begin{document}
\begin{titlepage}

\begin{flushright}
IFT-UAM-97-5\\
hep-th/9712136 \\
\end{flushright}

\vspace{3cm}
\begin{center}
\large{{\bf NEWTONIAN M(ATRIX) COSMOLOGY} } \\
\vspace{1.0cm}

\large{\bf {Enrique \'Alvarez and Patrick Meessen}} \\
\vspace{5mm}

Instituto de F\y sica Te\'orica, CXVI \\
and\\
Departamento de F\y sica Te\'orica,CXI\\
Universidad Aut\'onoma, 28049 Madrid, Spain \\
enrial,meessen@daniel.ft.uam.es
\end{center}
\vspace{1.5cm}

\begin{abstract}
 
A Newtonian matrix cosmology, corresponding to the Banks,
Fischler, Shenker and
Susskind  model of
eleven-dimensional M-theory in the infinite momentum frame as a supersymmetric
(0+1) M(atrix) model is constructed. Interesting new results are obtained, such
as the existence of (much sought for in the past) {\it static solutions}.
The possible interpretation of the off-diagonal entries as a background geometry
is also briefly discussed.

\end{abstract}
\vspace{15mm}

\noindent

\null
\end{titlepage}

\setcounter{footnote}{0}

\section{}
One of the most interesting recent results in string theory \cite{witten}
is that the strong coupling limit of type IIA strings is some unknown eleven 
dimensional theory, whose low energy limit is N=1
supergravity in 11 dimensions.
\par
Although this result, combined with the general framework of string dualities,
allows renewed hopes on the dream of unification, the fact that the eleven 
dimensional M-Theory does not have any known moduli associated to a coupling 
constant makes further progress  difficult.
\par
One of the few attempts to describe in a concrete way M- Theory is the M(atrix)
model of Banks, Fischler, Shenker and Susskind \cite{bfss,bilal}. 
The starting point is the seminal idea of \cite{ew} on the general 
description of  a
system of a given number, N, of D 
0-branes by means of the dimensional reduction of super-Yang-Mills in (1+9)
dimensions to (1+0) dimensions, i.e. a supersymmetric quantum mechanical model
with bosonic degrees of freedom given by nine N by N (hermitian) matrices
in the Lie algebra of U(N), $X^i$. To be specific,
the bosonic part of the action is:
\be
S =  \int dt \quad tr ( \frac{1}{2 R_c}(D_0 X^i)^2 + {R_c\over 4 l_{11}^6}
([X^i,X^j])^2
\ee
where $R_c$ is the proper size of the spacelike circle (becoming lightlike
when $\epsilon \downarrow 0$; this is a technical point discussed
by Bigatti and Susskind in \cite{bilal}) and $l_{11}$ is the scale set by the 
eleven-dimensional Newton's constant, $G_{11}\equiv l_{11}^9$.

The bold hypothesis of BFSS is that {\it precisely} this lagrangian (when 
supersymmetrized) describes the 11-dimensional dynamics of M-theory, albeit
in a frame boosted in the $x^{11}$ direction, the infinite momentum frame, IMF.
(widely used to describe some aspects of the physics of partons in the 
past \cite{ks}). It has been proven
that it reproduces low energy supergraviton scattering \cite{bb},
 and also membranes, 
following earlier suggestions in \cite{dwln}.
In order to be able to
consider arbitrary sets of particles, it seems necessary to take the limit
$N \rightarrow \infty$.  
\par
It is not clear what precisely the physical meaning of the fact that D 
0-branes are described by ``coordinates-matrices" is. On the one hand,
it is clear that the eigenvalues of the matrices
(wich will be denoted throughout
by $y^i_a$, where $a = 1,\ldots,N$ ) label the ordinary positions, 
whereas the off-
diagonal terms (denoted generically by $z^i_{ab}$) are related to the 
interactions between the particles
themselves\footnote{This position seems to be
at variance with the one expressed in \cite{bfss}, in which the relative distance
between two separate ``clusters'' in a block-diagonal matrix, with  blocks 
$Y_a$ ($N_a \times N_a$) and $Y_b$ ($N_b \times N_b$), with 
$N= N_a + N_b$ is $R_{a,b} \equiv
|\frac{tr Y_a}{N_a} - \frac{tr Y_b}{N_b}|$, but it is nevertheless the simplest
way of getting {\it gauge invariant} expressions for the coordinates themselves,
a very desirable property from our point of view.}.
It could be that Nature 
is described at some fundamental level
by non-conmutative geometry, and that non-separability is built in the theory
at a fundamental level. 

\par

At any rate, it seems clear that the model captures at least some relevant
aspects of interesting eleven dimensional physics.

\section{}
In the IMF relativistic invariance is hidden, and only galilean one
is manifest. Energies are given  up to a constant by  
$ H = \frac{k_{\bot}^2}{2\eta}$, where the ``Newtonian'' mass is related to
the eleventh component of the momentum in the IMF, $k_{11}\equiv \frac{\eta}{
\sqrt{2}}e^{\omega}$, where $\omega$ is the hyperbolic rotation angle 
determining the boost ($\omega\rightarrow\infty$). The energy 
in the unboosted frame is clearly given by
$p_0 \equiv \frac{1}{\sqrt{2}}(\eta + \frac{k_{\bot}^2}{2\eta})$,
and the unboosted eleventh component of the momentum is $p_{11}\equiv 
\frac{1}{\sqrt{2}}(\eta - \frac{k_{\bot}^2}{2\eta})$. Another point of view, 
pioneered by Susskind, and briefly alluded to in the prevoius section,
 starts from discrete light cone quantization, in which
the light-like coordinate $x^{-}$ is compactified to a circle of circumference
$2\pi R$. This means that the spectrum of $p_{-}(\equiv \eta)$ is discretized:
 $p_{-} = N/R$,
and we assume that $N$ (to be identified later on with the number of partons) 
is a non-negative integer. Eventually the limit $N\rightarrow \infty$ is 
necessary.
\par
This means that if we want to study cosmology in this frame (the purpose of
the present paper), it would
be some form of Newtonian cosmology. It could be thought premature 
to speculate in this direction, but in our opinion it is always useful
to be able to imagine a cosmological scenario compatible with the
best candidate to date for a fundamental theory\footnote{Some previous attempts
in the broad framework of Kaluza-Klein cosmology are to be 
found in \cite{freund}.}.
\par
Newtonian cosmology is a beautiful theory. It is almost a deductive one;
starting from the {\it cosmological principle} (postulating that the universe
presents the same aspect from every point except for local irregularities)
and assuming further that there is a universal force ({\it the cosmological 
constant}) proportional to the distance, all results of the theory flow
 smoothly \cite{bondi}.
\par
As has been stressed by Bondi, the main subject is the study of the motion 
of the substratum, which for our purposes will be idealized as the streaming
of a uniform fluid. The concrete implementation of the cosmological principle
will stem from the assumption that to all comoving observers the model presents
the same appearance (at a given {\it Newtonian} time).
\par
Although this is the standard formulation of the cosmological principle, given
the fact that the object of interest in the present work is a Newtonian theory
 with {\it extra dimensions}, some modifications can be thought of.
It could be applied to the ordinary three spatial dimensions only, for
example, which would constitute a sort of ``broken symmetry'' phase. From 
this point of view, the present work is restricted to
the {\it symmetric phase}.
(Non-symmetric phases would be related to the study of {\it cosmological}
(time-dependent)
solutions of ordinary supersymmetric Yang-Mills field theory (in n+1
 dimensions), at least for 
toroidal $\mathbb{T}^n$ compactifications, according to the standard 
procedure\cite{taylor}). Another window this sort of considerations would open up
is the cosmological study of the effects of T-duality \cite{gpr}
in this context.
\par
The cosmological principle implies striking
consequences on the allowed velocity
field for the substratum. Let us consider two {\it fundamental} (that is,
comoving with the substratum) observers, $O$ and $O^{\prime}$,
 measuring the velocity of a given point
$P$. Observer $O$ will report $V^i (X^j)$, and observer $O^{\prime}$
will report $V^{\prime}{}^i(X^{\prime}{}^j)$. Uniqueness of the velocity at 
$P$ (plus the Newtonian law of addition of velocities) 
then necessarily implies that $V^{\prime}{}^i( X^{\prime}{}^j) = 
V^i (X^j) - V^i( X^j - X^{\prime}{}^j )$
Now, the cosmological principle imposes $V^{\prime}(x) \equiv V(x)$, which 
means that the velocity field must be linear in the coordinate labels:
$V^i (X^j) = A X^i B$, where $A$ and $B$ are $N\times N$,
possibly, time-dependent matrices.
\par
We are assuming here that the substratum, is constituted somehow of D-0 branes,
in a regime in which a matrix description of them is compulsory. A fascinating
question, which has been left out for future work, is the construction
of a dual, complementary description in terms of membranes.
\par

In the standard {\it scalar} case, isotropy further requires that there 
is a single
function of time, $f(t)$ such that $\vec{v} = f(t) \vec{r}$. If we want this
 property to be preserved on the eigenvalues of the coordinate matrix $X$, and
further demand hermiticity, we are left with
\be
V^i(X^j) \equiv f(t) S  X^i  S^{+}
\ee
where $S$ is a unitary matrix. In the scalar case, this equation can be further
 integrated to $\vec{r} = R(t) \vec{r}_0$, with $f \equiv \dot{R}/R$. A 
similar Ansatz
in the present situation would be
\be
X^i\equiv R(t) U X^i_0 U^{+}
\ee
again, with a unitary $U$, yielding
\be
\dot{X}^i = \frac{\dot{R}}{R} X^i + R U [ U^{- 1} \dot{U}, X^i_0] U^{- 1}
\ee
\par
Assuming that the stress-energy from all matter of the Universe is 
dynamically negligible\footnote{This assumption could easily be bypassed},
the only remaining force is the {\it cosmological constant}
$F^i = \lambda X^i$,  (introduced in the newtonian context by Neumann and 
Seeliger in 1896 \cite{bondi}) which means that the dynamical 
Newtonian equations following
from the BFSS action principle (after a trivial rescaling) are
\be
\ddot{X}^i +  \sum_{j} [X^j,[X^j,X^i]] = \lambda X^i
\ee

\par
The preceding equations convey an interesting generalization of the standard vacuum
equations of scalar Newtonian cosmology. The {\it eigenvalues} of the 
coordinate-matrices obey {\it exactly} the standard equations as in \cite{bondi}.
(Nevertheless, if the unitary matrix $U$ is non-trivial the labels of the 
eigenvalues suffer a time-dependent permutation).
The most conservative interpretation of the other entries of the matrices is 
that they are related to the interactions. More on this later.
\par
Although it is not the purpose of the present work to be exhaustive about it,
 there are very interesting new solutions. One of the simplest stems from the 
assumption that the unitary matrix $U = 1$, leading to
\be
\ddot{R} X_0^i +  R^3 \sum_j [X_0^j , [X_0^j,X_0^i]] = \lambda R X_0^i
\ee
The trace of the above is
\be
(\ddot{R} - \lambda R) tr X_0^i = 0
\ee
There are then two possibilities: either $tr X_0^i \neq 0$, in which case 
the standard
scalar result is recovered: $ R = R_0 \cosh (\sqrt{\lambda}(t - t_0))$
 (if $\lambda > 0$), or $R = R_0 \cos (\sqrt{- \lambda}(t - t_0))$ (if $\lambda
< 0$ ); or else $tr X_0^i = 0$, which means that $X_0^i$ can be expanded in 
generators of su(n): $ X_0^i \equiv \sum_a u_{ia} T_a $. To be specific,
assuming the simplest case $N = 2$, and $u_{ia} = \theta(3 - i) u_{ia}$,with 
furthermore $\sum_c u_{ic}u_{jc} = \frac{1}{2}\delta_{ij}$
(equivalent to choosing a set of 3 orthonormal ordinary three-vectors ),then the
$u_{ia}$ are subject to no further restriction, whereas R obeys:
\be
\ddot{R} -\lambda R +  R^3 = 0
\ee
whose solutions are, generically, given in terms of elliptic funcions.
\par

There is a fascinating particular solution, though, in the case $\lambda > 0$,
and it is a {\it static} one, $R = \sqrt{\lambda}$.
This is very curious\footnote{Although perhaps not
so surprising, given the fact
that our building blocks for the substratum are D-0 branes,
which are well known
to be BPS states, and, as such, able to survive in static configurations
of equilibrium}, because (scalar) Newtonian Cosmology was 
abandoned essentially
since the first attempts in the nineteenth century
because it does {\it not} allow static solutions, and this
was exactly everybody's prejudice until Milne and MacCrea
revitalized the subject
(in 1934) drawing on insights from relativistic cosmology as well as on 
Hubble's data 
on the expansion of the Universe.

Apart from the static solution, $R=\, constant$, we can multiply the 
above equation with $\dot{R}$ and find
\begin{equation}
    \dot{R}^{2}\,-\, \lambda R^{2} + \frac{1}{2} R^{4} \;=\; {\cal C} \; ,
\end{equation}
where $\cal C$ is some constant. One one can follow \cite{bondi} in his 
analysis of solutions, but for the sake of briefness let's give the 
explicit expression for the case $\lambda =0$: Integration imposes that
${\cal C}\geq 0$, and we find that the 
solution is given by \cite{abra}
\begin{equation}
  R(t) \;=\; 
     a\,\mbox{\bf sd}\left(\sqrt{2}a(t-t_{0})\left| \frac{1}{2}\right.\right) \; , 
\end{equation}
where $a^{4} = {\cal C}$.
\par
The above solution can be easily generalized to arbitrary
$N$. Note that the solution is based on the asumption that 
we have matrices $X_{0}$ such that
\begin{equation}
  \sum_{j} \, [ X_{0}^{j}, [X_{0}^{j},X_{0}^{i}]] \;\equiv\; X_{0}^{i} \, .
\end{equation}
If we now decompose the $X$'s on a base of $su(N)$, using the Lie-data 
$\{ T_{a}, {f_{ab}}^{c} \}$, we see that the above equation reads
\begin{equation}
 X^{ia}\,\sum_{j}\, X^{jb}X^{jc} \, {f_{ab}}^{d} f_{edc} \;=\;
 {X^{i}}_{e} \; ,
\end{equation}
where we have used the Killing metric on $su(N)$, $g_{ab}$ say, 
to lower the index.
It is then easy to see that iff we take the $X^{jb}$ such that, reverting
to vector notation,
\begin{equation}
  \sum_{j} \vec{X}^{j}\, \left(\vec{X}^{j}\right)^{T} \;=\; g \; ,
\end{equation}
we automatically satisfy the above equation, due to the definition of the
Killing metric. Note that we are using Hermitean generators, so that the
Killing metric is positive definit.
\par

A completely different solution is obtained for one-dimensional motion (still 
with N=2, for simplicity). A rotation
can always be made so that $X_0 = \sigma_3$. Making the general ansatz
$U = u^0 + i \vec{u}\vec{\sigma}$, it follows easily that
\begin{displaymath}
X = R(t)
\left(\begin{array}{cc}
1 - 2 \sin^2 \theta(t)&- 2 \sin \theta(t) \cos\theta(t)\, e^{i\phi (t)}\\[.1cm]
- 2 \sin\theta(t)\cos\theta(t)\, e^{-i\phi (t)} & 2\sin^2 \theta(t) - 1\\
\end{array}\right)
\end{displaymath}
The simplest case would be $\theta = \pi/4$, giving
\be
\ddot{R} - R \dot{\phi}^2 - \lambda R = 0
\ee
\be
R \ddot{\phi} + 2 \dot{R}\dot{\phi} = 0
\ee
Again, a  curious thing about these equations is that they allow {\it static} 
solutions
when $\lambda < 0$; $\phi = \sqrt{-\lambda} t + \phi_0$ and $R = R_0$ arbitrary.
\par
Given the fact that the matrix $U$ is not trivial anymore, there is now a 
time-dependent
permutation of the labels of the two fundamental particles in our 
oversimplified model.
\section{}
The problem of the physical interpretation of the off-diagonal terms must
now be tackled. The sort of solutions presented here (at least when $\lambda = 0$)
correspond to particles as free as they can be in the model. The most 
conservative approach would be to identify a spacetime metric such that 
their geodesics coincide with the trajectories of the eigenvalues in M(atrix) 
Cosmology. There are problems, however. The hamiltonian of the M(atrix) 
model yields an expression for the 
energy,
which, when the effects on the eigenvalues $y_a^i$ are separated from the
rest ($z_{ab}^i$) is , in a somewhat symbolic notation,
\be
E = \frac{1}{2}\sum_{a=1}^{N}(\dot{y}_a^i)^2 + g^{ij}_{ab}\dot{z}^i_a\dot{z}^j_b
+ V(y^i_a,z^j_{bc})
\ee
where the $ g^{ij}_{ab}$ are quadratic functions of the $y^i_a$. This means 
that the classical equations of motion constitute a coupled system of $N^2$
differential equations of the type:
\be
\ddot{y}^i_a = F^i_a(y^j_b,\dot{z}^k_{cd},z^l_{ef})
\ee
where the functions $z^i_{ab}(t)$ obey:
\be
\ddot{z}^i_{ab}=G^i_{ab}(y^i_a,\dot{y}^j_b,y^k_c,z^l_{de})
\ee
\par
In the limit $N\rightarrow\infty$ we can assume the we have an $\infty^{9}$
of {\it fundamental observers}, $a\equiv a^i \in \mathbb{R}^{9}$, such that
for each point there is a unique trajectory passing through it. There are many
curved spaces such that the trajectories $y^i_a = y^i_a (t)$ are geodesics
of it; it suffices to take the congruence of fundamental observers as timelike lines;
parametrized by $ x^i = a^i$, (so that the change of coordinates is defined by
$x^i = a^i (y^k,t)$) and  a metric, for example such as
\be
ds^2 = dt^2 - R(t)^2 \delta_{ij} dx^i dx^j
\ee
(which reduces to $ds^2 = dt^2$ on the trajectories by using $\frac{\partial a^i}
{\partial y^k} \dot{y}^k + \dot{a}^j = 0$).
\par 
In the initial coordinates it reads:
\be
ds^2 = (1 - R^2 \dot{a}^2)dt^2 - R^2\delta_{ij}\frac{\partial a^i}{\partial y^k}
\frac{\partial a^j}{\partial y^l}dy^k dy^l - 2 R^2 \delta_{ij}\frac{\partial a^i}{\partial y^k}\frac{\partial a^j}{\partial t}dy^k dt
\ee

There is a certain latitude as to how to choose the function $R(t)$. The simplest way
 would be to demand the we recover newtonian cosmology on the constant time 
hypersurfaces (that is, on the 9-space orthogonal to the trajectories of our 
fundamental observers). This leads to its identification with the same function 
$R(t)$ of preceding paragraphs.

\par
Neither should constitute a big surprise the fact that the above metric
is not flat in general. D-0-branes only feel the metric and the RR-vector,
so that the background constructed as above  need, a priori, not
satisfy the IIA-Sugra equations. A legitimate question is whether, building on the 
data obtained from M(atrix), we can find a 
IIA-Sugra solution a posteriori.
\par
Another point is that this technique gives naturally a 10-dimensional background.
It could be thought more natural to interpret it as an 11-dimensional one, but
it does not appear easy in our framework (other than definining 11-dimensional 
decompactifications of the 10-dimensional manifold M).

\section{}
Quantum effects in the IMF are usually simpler to compute, because vacuum effects
are absent. This is due to the fact that as the boost parameter $\omega
\rightarrow \infty$ those processes corresponding to diagrams 
which have internal lines carrying negative $\eta$ will go to zero, and only
those with positive $\eta$ survive. Quantum cosmology would then be 
particularly transparent in this frame, in which the physical interpretation 
of the wave function
of the universe could follow the guidelines of hadronic wavefunction in parton 
physics. We hope to be able to return to this topic in the future.

\par 

It is perhaps worth stressing, to conclude, that the set of ideas here introduced
with a motivation based on  M-Theory, could also be contemplated as a natural
generalization of Newtonian Cosmology {\it per se}.  The basic equations could
easily have been derived just after the introduction of the matrix concept
by Sylvester around 1850.

\section*{Acknowledgements}
We have benefited from many discussions with A. Casas, C. G\'omez and
 T. Ort\y n. This work has been supported by EU TMR contracts ERBFMRXCT960012
and ERBFMBICT960616, and CICYT grants AEN/96/1655 and AEN/96/1664.
%

%
%
\end{document}